\documentclass[preprint,12pt]{elsarticle} 
\usepackage{graphicx} 
\usepackage{amssymb} 
\usepackage{amsmath} 

\journal{Nuclear Physics A} 

\begin{document} 

\begin{frontmatter} 

\title{Short History of Nuclear Many-Body Problem}

\author[a]{H. S. K\"ohler}
\address[a]{Physics Department, University of Arizona, Tucson, 
Arizona 85721}

\begin{abstract} 
This is a very short presentation regarding developments in the theory of nuclear
many-body problems, as seen and experienced by the author during the past 60 years
with particular emphasis on the contributions of Gerry Brown and his research-group.
Much of his work was based on Brueckner's formulation of the nuclear many-body
problem. It is reviewed briefly together with the Moszkowski-Scott separation 
method that was an important part of his early work. The core-polarisation 
and  his work related to effective interactions in general are also addressed.   
\end{abstract}

\end{frontmatter} 

\section{Dedication} 
\label{sec:ded} 

The main content of this work was presented at the Gerry Brown memorial 
conference  in Stony Brook. 
I first met Gerry at least 55 years ago while he was with Rudolf Peierls in
Birmingham and I was a graduate student in Uppsala and at CERN. 
I never worked directly with him 
(although he tried to recruit me for some of his projects). But I did
interact with him in various ways over the years.  In 1959 he was my 
opponent at my PhD thesis \footnote{On optical model with spin-orbit coupling}  
defense in Uppsala (Sweden). 
Last time we met was at Eyvind Osnes retirement conference in Oslo in 2008. He told 
me then, after my talk,  that I should have "spruced it up" like he himself
does.  He was a good friend.
Gerry had many collaborators not only among his many students. He was always
able to make others interested in problems he considered important. It was
one of his strengths. That together with his enthusiasm, physical insight and
intuitive thinking is how he will be remembered.
His contributions to the problems of nuclear physics were dominating and
will be ever-lasting. It is not possible to cover more than a small fraction of 
his work on many-body physics in this short talk.

\newpage 

\section{Introduction} 
\label{sec:intro} 
I find the history of physics (almost) as fascinating as physics itself.
A historical perspective shows  a scenario of  ideas and people behind the ideas
often not found in published papers. The real physics is of course found in what
experiments reveal to us. The human brain seeks to understand these physical
phenomena and that is what theorists are trying. It is a fascinating
interaction between us and the world around us. Theory is the subject of
this presentation.

When does the history of the nuclear many-body problem start? One of the
great discoveries was the nuclear shell-model. Liquid drop, collectivity,  was
the predominant and succesful picture theorists had before that. So how can one 
explain the success of the seemingly contradictive picture that the
shell-model presented? Another problem: nuclear saturation. A possible
explanation: N-N interactions are repulsive at short distances. \cite{jas51}. But
how can one reconcile the strong interactions with a shell-model? How can one
deal with the strong and even infinitely repulsive forces computationally?
These were some of the nuclear physics problems some 60 years ago.

The stage was set for someone to come up with a many-body theory of nuclear
structure. The first successful nuclear many-body theory was that of K.A.
Brueckner's. Gerry Brown's (and others) nuclear strucure work was based on
this theory. 

In the theoretical treatment of the nuclear many-body problem ,
nuclei are in general  considered to be composed of nucleons (protons and neutrons)
interacting with some specified forces without internal degrees of freedom.  
The solution of this problem is hampered by two difficulties.\\ 
I. The strength and complexity of  interaction(s) that are also 
unknown in details. \\
II. The mathematics to solve for  the physical properties of a many-body
system. \\
The first problem (I.) is still being worked on in several ways.
A method to overcome the second problem (II.) was presented by K.A. Brueckner some
60 years ago.  It showed how the strong interaction can be replaced by
an 'effective' softer interaction, more manageable to handle. 
The invention of the shell-model led to a very active research to interprete the
experiments on nuclear spectroscopy.
A crucial part of this work was of course then the choice of NN-interaction. 
Gerry Brown, understanding the significance of Brueckner's
work, commenced 
very successful shell-model calculations together with collaborators 
using the Brueckner 'effective' interaction, the reaction-matrix, to do the job.

\section{Brueckner theory}
Related to the shell-model is the optical model of the 50's, that pictures
nucleons moving in a mean field. It was explained by Francis and Watson 
as a multiple scattering problem with elementary scatterings being 
via 'soft'  T-matrices
rather than 'hard' N-N bare interactions.\cite{fra53}
This idea was picked up by Brueckner: Maybe a nuclear many-body theory for
bound states could also be built on T-matrices instead of the, at the time, 
conventional efforts using NN-interactions with slowly converging or diverging 
results.
But the T-matrix is complex
\begin{equation} 
T=v+v\frac{1}{k^{2}-k'^{2}+i\eta}T \sim e^{i\delta}sin({\delta})
\label{T} 
\end{equation} 
with the $\sim$ indicating the complex diagonal element. For calculating the
real binding energies of nuclei 
it seemed to make some sense to replace this complex $T$-matrix with the 
real Reactance matrix (the R-matrix)  defined by a principal value
integration.\cite{bru54}
\begin{equation}
R\sim tan(\delta)
\label{R}
\end{equation}
replacing the bare interaction with an 'effective' interaction
\begin{equation}
V(k)\sim tan(\delta(k))
\end{equation}
This idea had some degree of success when calculating nuclear bindings.
BUT, the R-matrix refers to a scattering problem with boundary problems
different from that of a bound state. It is fairly easy to show that if putting
two interacting particles in a box, square or Harmonic
oscillator (Busch formula) the
binding energy is not $\sim tan(\delta)$ but rather just $\delta$.
\footnote{The Busch formula expresses the binding energy of two nucleons in
an oscillator well in terms of phase-shifts. I recently showed that the
SHIFT in energy like for the square box is given by $\delta$}
In the scattering problem one deals with a continuum set of states but in
the bound state problem with a discrete set of states.
"Infinite" nuclear matter still implies a bound state problem. Summation
over a discrete set of states no matter how dense is different from
integration over a continuum.\cite{fuk56,dew56,rie56}
\footnote{The difference between $\delta$ and $tan(\delta)$ may of course be negligible for
small $\delta$. With large scattering lengths especially with $\delta=\frac{\pi}{2}$
it makes a big difference.}
Brueckner later improved the theory to include medium
effects. 
The $\delta$-(or phaseshift) approximation is however still good
at low density and also for 'weak' interactions as well as
for large angular momenta $l\geq 4$, when
medium-effects can be neglected.

What about the medium, many-body effects? We deal with an assembly of
nucleons, a fermion-system. When summing over intermediate states, 
the occupied states  should therefore be excluded,  modifying the
$T$-matrix (\ref{T}). Brueckner together with
Wada defined this modified effective interaction, the reaction matrix $K$
\cite{bru56} to get
\begin{equation}
K=v+v\frac{Q}{k^{2}-k'{2}}K
\label{K0}
\end{equation}
with $Q$ being a Pauli-operator.
Note that $K$ is real. No summation (integration) over a pole. No longer a
discrete-continuum controversy.
But the shell-model says that the nucleons move in a mean field $U(k)$. Accordingly
energies are not $e(k)=k^{2}$ but rather $e(k)=k^{2}+U(k)$.  

This additional modification results in what is usually referred to as 
Brueckner theory, summarised as
follows.
Brueckner Reaction Matrix $K$ defined by \cite{bru58}:
\begin{equation}
K=v\frac{Q}{e(k)-e(k')}K
\label{K}
\end{equation}
Total energy (first order):
\begin{equation}
E_T=\sum k^2+\frac{1}{2}\sum K
\label{ET}
\end{equation}
Mean field:
\begin{equation}
U(k)=\sum K
\label{Uk}
\end{equation}
This equation for $U(k)$ is often referred to as a \it Brueckner  self-consistency
\rm because $U$ is a functional of $K$.

The Brueckner $K$-matrix sums ladder and mean-field propagations to all orders.
Infinite nuclear matter calculations show saturation and binding energy in
remarkable agreement with experimental information.
Important physics is included in this first order in $K$ approximation. Improved
results can (in principle) be obtained by higher orders but convergence problems
are not well understood.\cite{bet71,hsk75}
\begin{figure}
\begin{center} 
\includegraphics[width=0.32\textwidth,angle=-90]{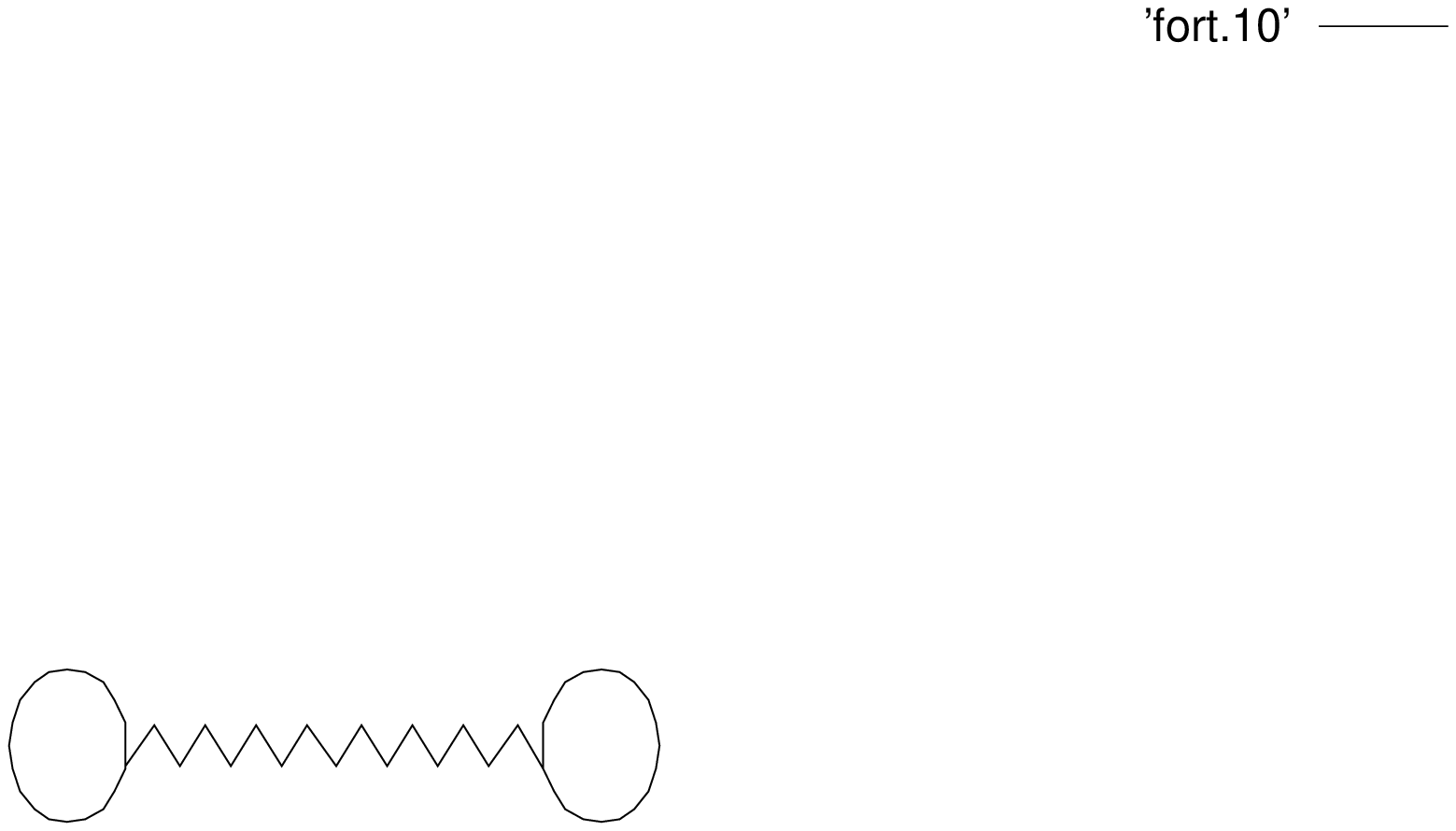} 
\includegraphics[width=0.32\textwidth,angle=-90]{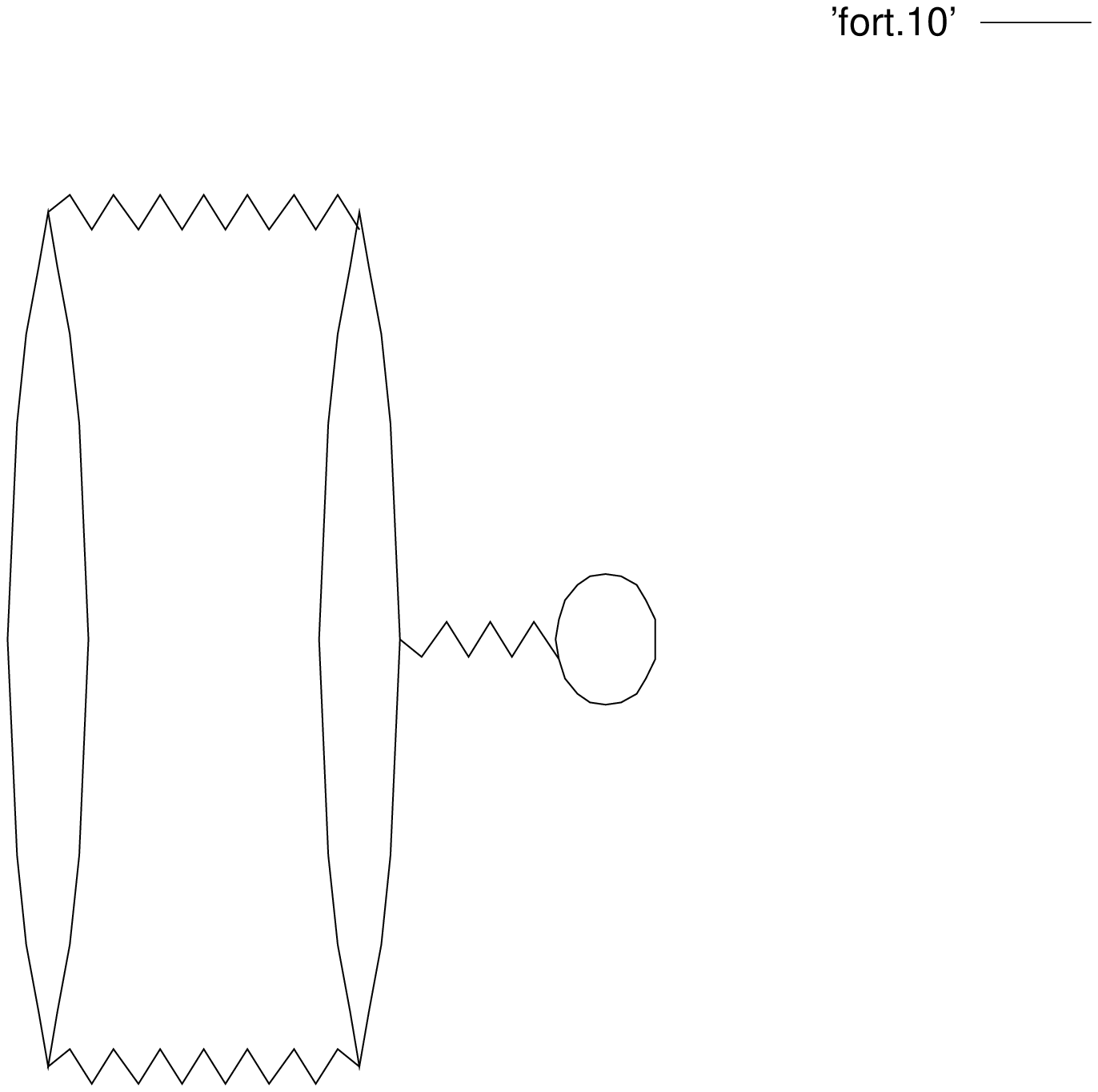} 
\caption{
Diagrams included in the Brueckner $K$-matrix approximation.
The diagram to the right is a three-body term. The interaction between two
particles is influenced by the presence of a third, the tadpole. This diagram
relates to the dispersion-term that is an important term in the
separation-method. See section below.
}

\label{Kfig} 
\end{center} 
\end{figure}

It is a zero spectral-width approximation. Spectral widths can be included in higher
orders  or Green's function calculations but showing relatively small corrections.

Gerry Brown (with co-workers) had several publications related to nuclear matter.
In a 1964 paper,  Gerry Brown and coworkers investigated the effect of various
approximations and off-energy shell propagations.\cite{bro64}
In a 1982 paper with Hans Bethe the focus was on compressibility an important
matter in relation to supernovaae explosions.\cite{bet83}

\section{Separation Method}
Much of the computational work on Brueckner theory in the early 60's was to find
suitable approximations as the access to computers at that time was limited.
The equations are nowadays very simply solved numerically.
The separation method introduced by Moszkowski and Scott \cite{mos60} for 
evaluating the reaction matrix proved   valuable not only as a computational 
tool but it also gives
valuable insight into the physics. It also relates to more recent ideas behind
EFT and $V_{low-k}$ as presented in a section below.
The method involves dividing the nucleon-nucleon potential $V$ into two parts
in ccordinate space, the
short-ranged part $V_s$ and the long-ranged part $V_L$. $V_s$ will have a repulsive
as well as an attractive part. A separation distance $d$ (in coordinate space) 
is chosen such that the scattering phase-shift for the potential $V_s$ 
is zero. This is of course momentum-dependent so that $d=d(k)$.
A slight modification of the method was made by K\"ohler \cite{hsk61}. He obtained
(in operator-form)
\begin{equation}
K=K_s+V_L+(\Omega-1)e(Q-1)(\Omega-1)+(\Omega-1)U(\Omega-1)+2V_L(\frac{Q}{e)}V_L
\label{sep}
\end{equation}
 $\Omega$ is the wave-operator for the short-ranged part therefore
related to correlations. Of particular significance is the fourth term
proportional to  the mean field $U(k)$  as well as the correlation-volume $I_D$
(the "wound-integral"). 
\begin{equation}
I_D=\int (\psi-1)^{2}d{\bf r}
\label{ID}
\end{equation}
where $\psi$ is the in-medium NN-correlated wave-function. (See Fig. (\ref{fig1})
below).

This fourth term is referred to as the \it dispersion term \rm. 
\begin {equation}
K_{disp}\sim I_D*U
\label{Kdisp}
\end{equation}
It is repulsive 
and  important for providing nuclear saturation. It is a 3-body term
with two particles  interacting with other particles in the
medium which is represented by the momentum-dependent mean field $U(k)$ .
It involves off-shell scattering and therefore model-dependent, while the
on-shell scattering is fixed by fits to free space scattering. 
The separation method was used by Gerry Brown and co-workers (and others) in
shell-model calculations.

\section{Core-polarisation}
\begin{figure}
\begin{center} 
\includegraphics[width=0.32\textwidth,angle=-90]{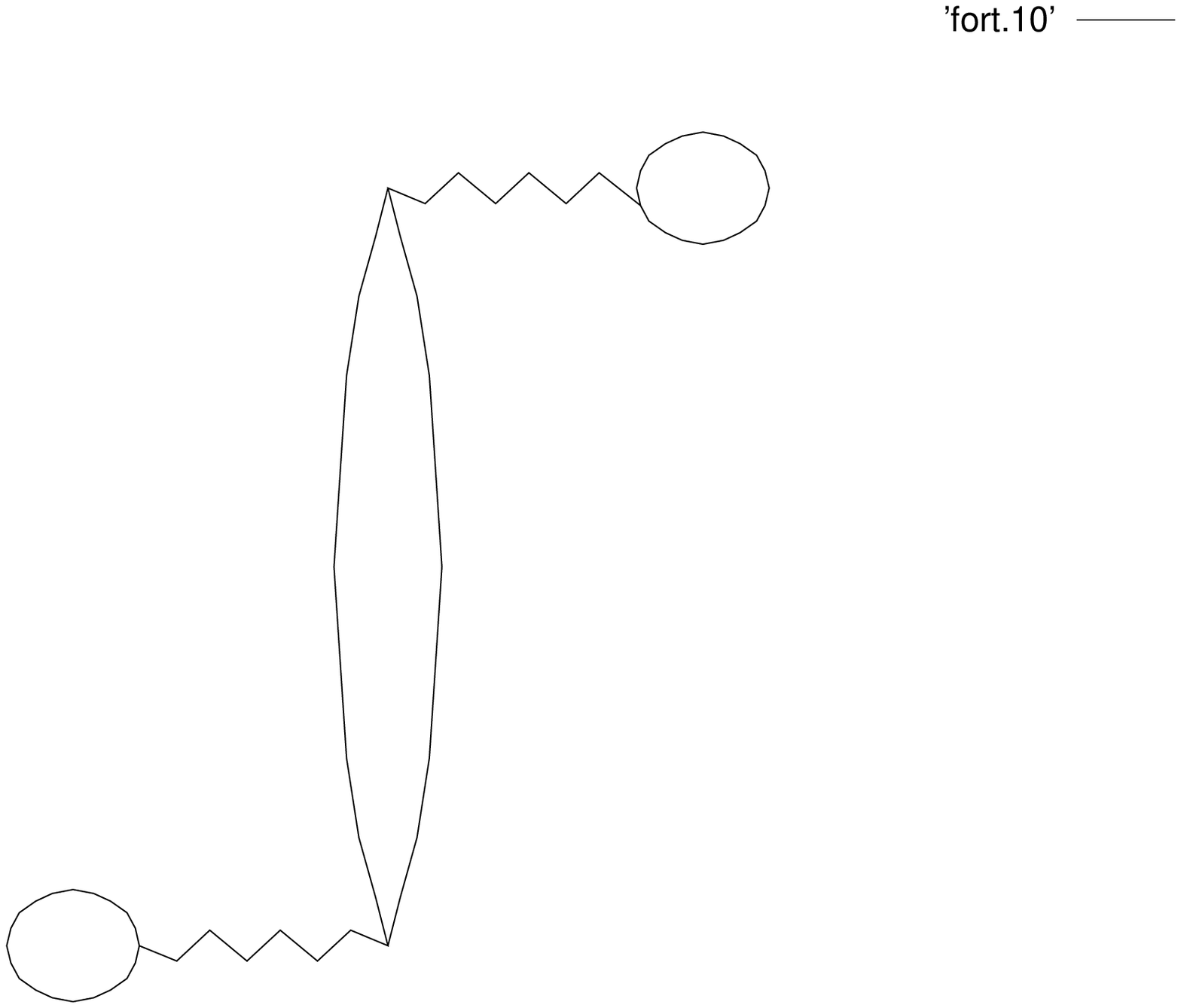} 
\includegraphics[width=0.32\textwidth,angle=-90]{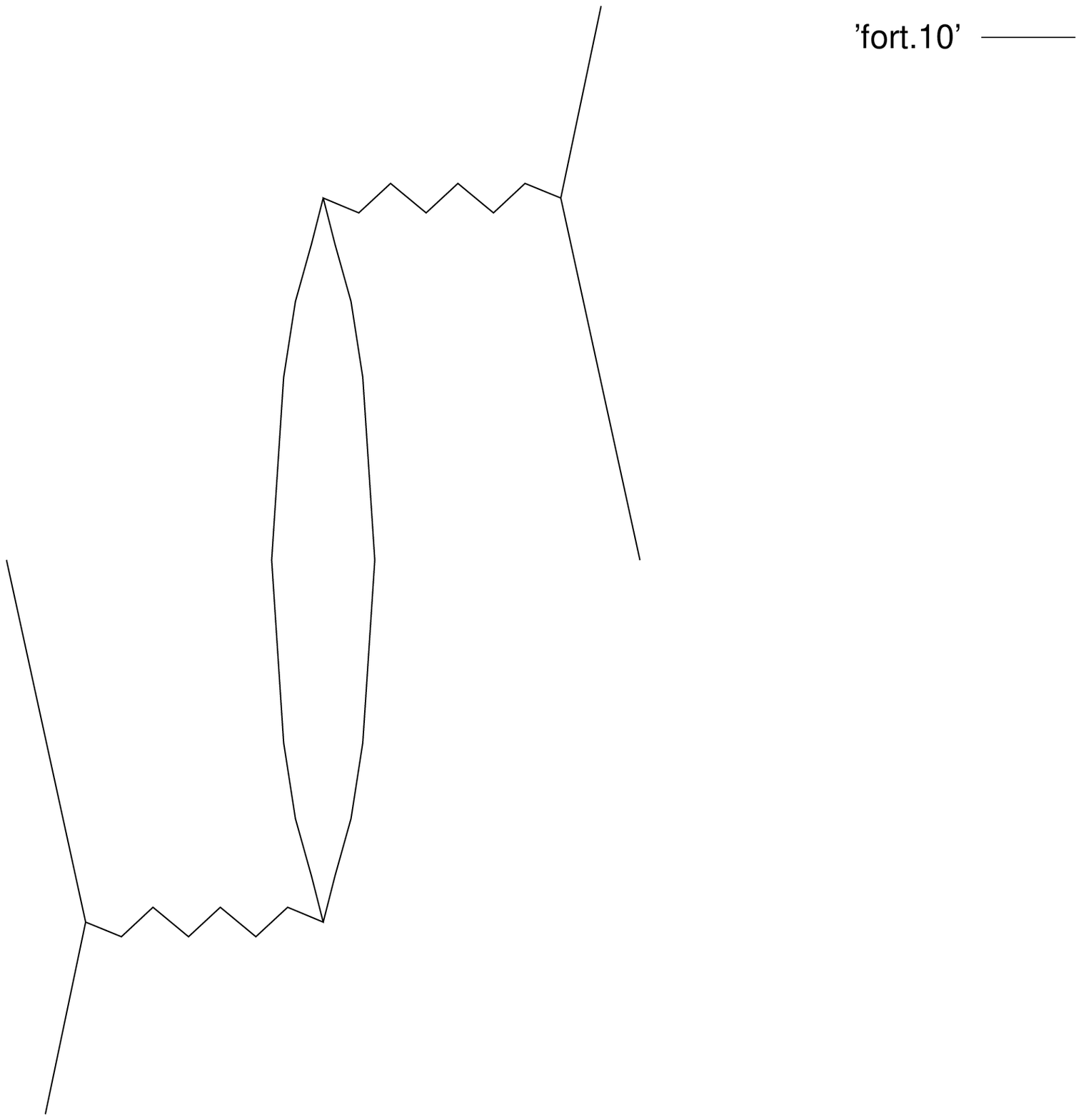} 
\caption{
Lowest order core-polarisation diagrams. 
To the left is the energy-diagram. Breaking the two
loops results in the core-polarisation diagram for the effective interaction
as shown to the right.
The energy diagram is cancelled by a proper choice of the mean field .
}

\label{fig0} 
\end{center} 
\end{figure}

While the Brueckner reaction-matrix  gives an
effective interaction based on 'realistic' forces between the constituent nucleons, 
results of early shell-model calculations  were not promising. Gerry Brown realised 
that an important
contribution to the  interaction, beyond the reaction-matrix, was missing. 
The clue to the problem  is that  particle-hole excitations are allowed in the finite
system. 
This results in an additional term , the \it core-polarisation \rm shown
diagramatically in Fig. (\ref{fig0})

It was first included in calculations  by George Bertsch \cite{ber65} for 
$^{18}O$ and $^{42}Sc$  using the Kallio-Kolltveidt force.\cite{kal64}
This paper was followed by the seminal papers by Tom Kuo and Gerry Brown on the
"Structure of Finite Nuclei and the Free Nucleon-Nucleon Interaction"
\cite{kuo66,kuo67,kuo68}. Here the reaction matrix as well as the 
core-polarisation was calculated from the Hamada-Johnston potential\cite{ham62} 
using the separation method.
The Kuo-Brown interaction has been, and still is used by many research-groups
in nuclear structure calculations.
\footnote{A more complete summary and history of these and following
works are found in the book by Gerry Brown and coworkers.\cite{bro10}}
Barrett and Kirson questioned the convergence of the core-polarisation diagrams.
\cite{bar70} In a paper by Anastasio et al the effect of the choice of shell-model
mean field potential is illustrated.\cite{ana76}

\section{$V_{low-k}$}
A many-body problem is always a two-part problem:\\
1. Interactions between particles e.g. 2-,3- etc interaction potentials.\\
2. A many-body theory. \\

The theory of nuclear forces has been a long-standing problem. (see e.g. ref.
\cite{mac89}). It is easy to construct NN-potentials that fit scattering
phase-shifts as well as deuteron data e.g. by inverse scattering and separable 
potentials.\cite{kwo97} But that is in general not enough. Off-shell scattering information is
needed for many-body calculations. This was for instance already emphasized by
Gerry Brown and coworkers in the 1964 paper. \cite{bro64} 
It was already stated above that the dispersion effect depends on off-shell
scattering.  This is actually a main 
problem in nuclear many-body calculations.  That is why we need a theory 
of nuclear forces going beyond the on-shell information obtained from scattering. 
And off-shell scattering is directly related to correlations, short-ranged as
well as tensor-induced.

It can however also be argued that the high energy component of the interaction
(i.e. the largely un-known short-ranged part) should be less relevant 
in dealing with low-energy
nuclear problems. The low and high energy components are separated when using the
separation method shown above with $V_L$ , the long-ranged part, being the
leading term. 

A different approach whereby the high energy component of the force was 
eliminated by integrating out the largely unknown and model dependent part of the
interaction was presented by Bogner et al \cite{bog01,bog02}. The resulting 
low-momentum
potential is referred to as $V_{low-k}$ and it is a function of the chosen cut-off
momentum $\Lambda$. This cut-off has a motivation similar to the cut-off in
coordinate space in the separation method. The near numerical equivalence was shown
in a paper by J. W. Holt and G. E. Brown.\cite{hol04}
 
A more fundamental approach to the separation of momentum scales is EFT (Effective
Field Theory) originated by Weinberg. \cite{wei90}
Not suprisingly, Gerry Brown was consulted by Weinberg.

\begin{figure}[!ht] 
\begin{center} 
\includegraphics[width=0.48\textwidth]{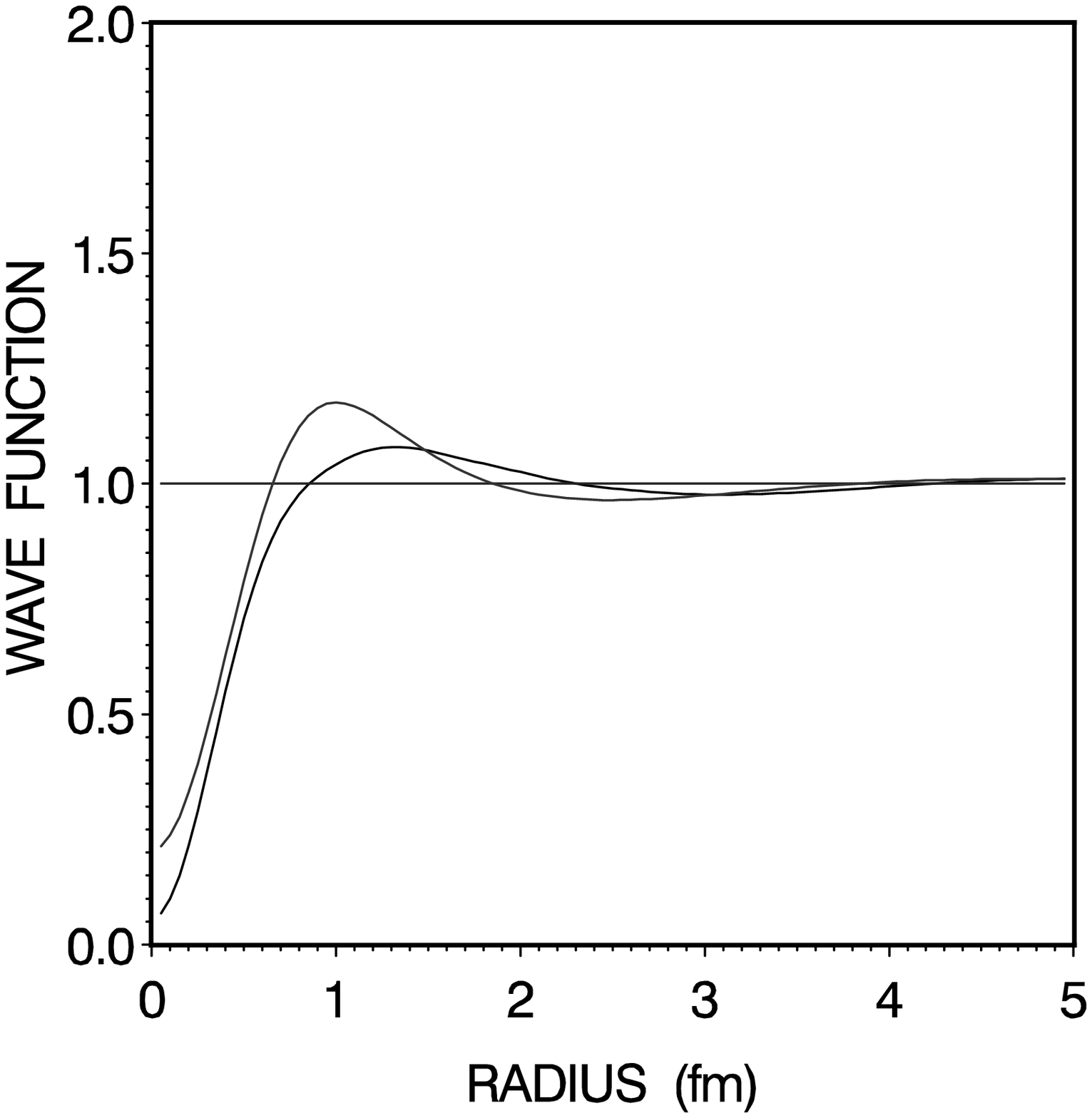} 
\includegraphics[width=0.48\textwidth]{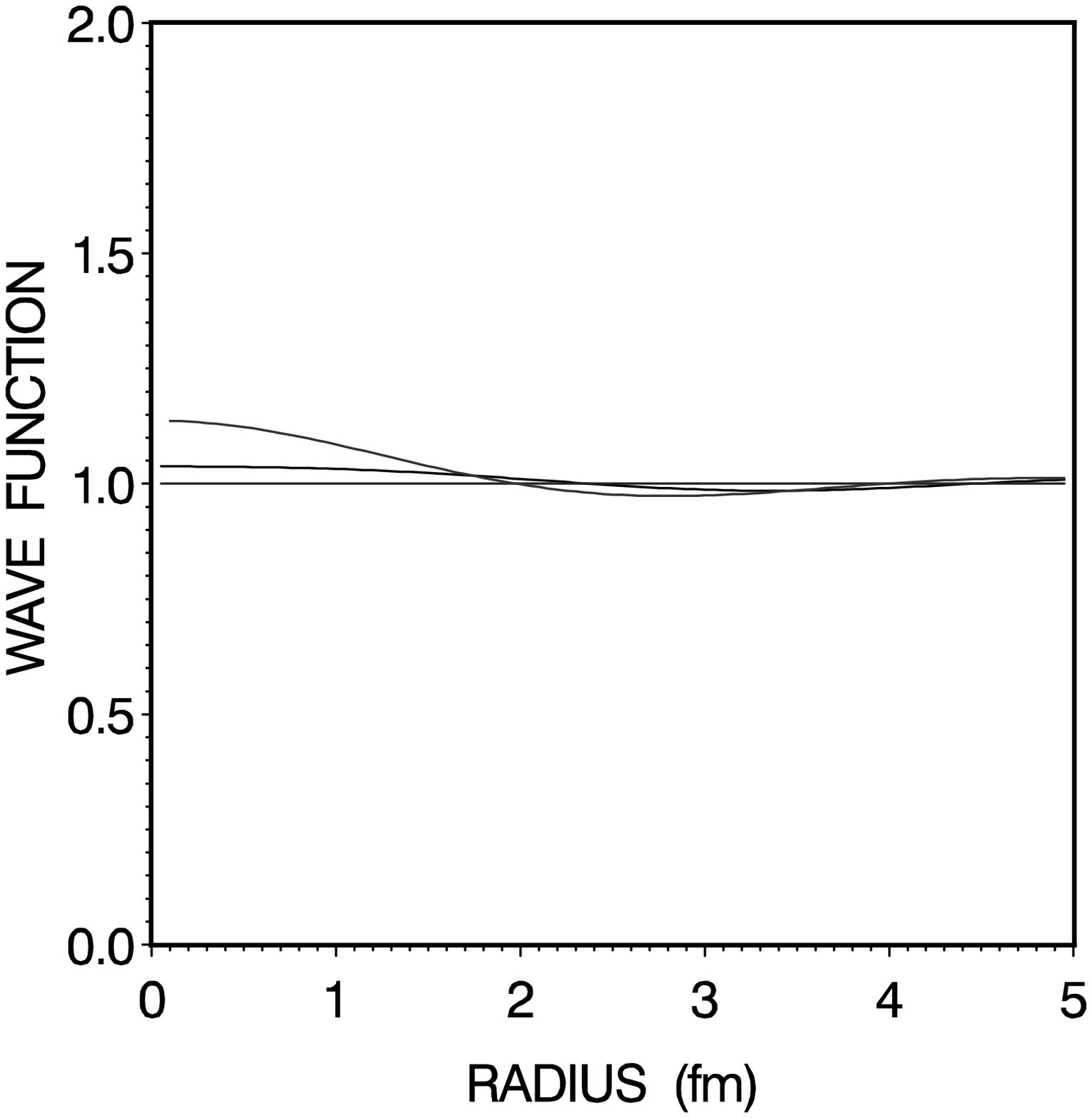} 
\caption{
Left curves:The correlated wave functions for triplet (upper curve) and for 
singlet interactions (lower curve) with a cut-off $\Lambda=9.8 fm^{-1}$. 
Right curves: Similar to the left curves but for a cut-off $\Lambda=2
fm^{-1}$.
}
\label{fig1} 
\end{center} 
\end{figure}

\begin{figure}[!ht] 
\begin{center} 
\includegraphics[width=0.48\textwidth]{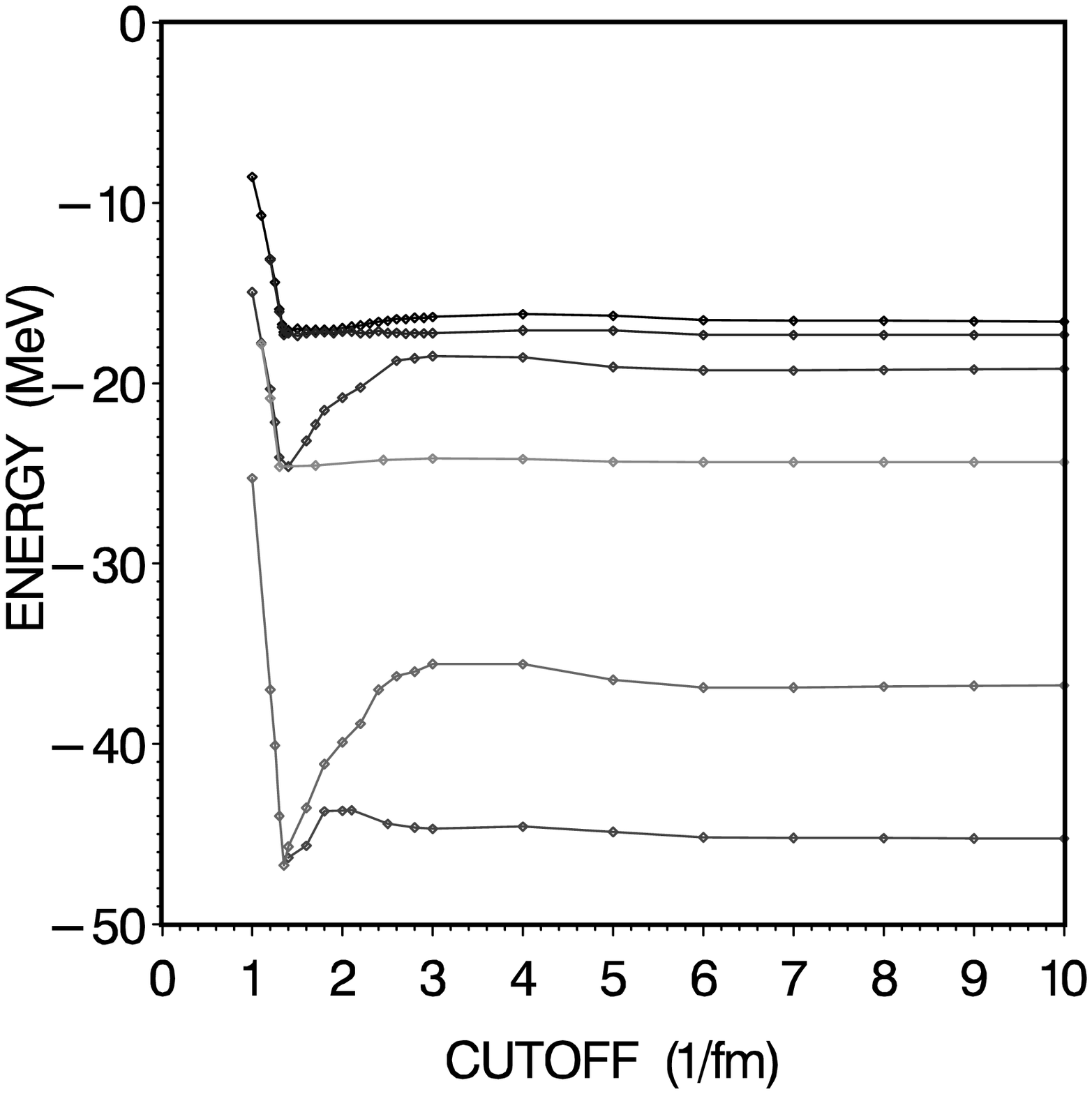} 
\caption{
The potential energy per particle is shown as a function of cut-off $\Lambda$.
The three set of curves show from top to bottom: singlet S-state, triplet S-state and
total contributions (all angular momenta). At each set there are two curves: 
The lower is without the mean
field (i.e. without dispersion effect) and the upper is with mean field, 
thus with less attractive effective force. The loss of correlations shows 
binding to increase dramatically for $\Lambda \leq 2.5$ .
Note also the difference in dispersion effect between the singlet and triplet
states. This agrees with the difference in correlations in the two cases as
shown in Fig. (\ref{fig1}).
}
\label{fig3} 
\end{center} 
\end{figure}

Correlations in nuclear matter \it are \rm however important. This is explicitly illustrated
in the separation method where the wound-integral (\ref{ID})  is a measure of
the correlations. The dispersion term (\ref{Kdisp}) , being  a
product of the mean field and the wound-integral , is  responsible for 
saturation and nuclear stability. A too small momentum cut-off quenches the
correlations so that $K_{disp} \rightarrow 0$ with no saturation.\cite{hsk07} The
situation is illustrated by the two set of curves in fig ~(\ref{fig1}) . 
Fig. ~(\ref{fig3})
shows the influence of the loss of correlations (and the dispersion-effect) 
on the binding energy. 
It should also be noted that the dispersion effect is much smaller in the
singlet channel in agreement with Fig. ~(\ref{fig1}). 
These are
results of standard Brueckner calculations at normal nuclear matter density.

\section{Summary and Future}
The last fifty years of developing methods to solve the nuclear many-body
problems have been an exciting time. Keith Brueckner's pioneering work and Gerry
Brown's application of Brueckner's method to the nuclear shell-model problem will
remain as corner-stones in the history of nuclear theory. 
The outcome of the numerical work relates of course closely to the (free)
nucleon-nucleon interactions. These are in particular complicated in relation to
induced correlations. The last few years have shed important light on this problem
assuming that the low energy nuclear problem should not be impacted (in a major
role) by short-ranged (high energy) components of the forces. Much care has however
to be taken in order not to loose important effects of tensor and short-ranged
effects. This was illustrated above and the relation to the dispersion term in the
separation method expansion was emphasized. This is a three-body term and has
effects similar to that of three-body forces but with separate physical content.

The nuclear many-body methods have
been of significance not only for developments in nuclear
theory, but also for
developments in atomic and molecular physics as well.
It has also inspired researchers to develop competing methods such as coupled
cluster, hypernetted chain, Green's function methods to name a few. 
The density functional methods should also be mentioned in this
context.
All of the improvements in applying these methods have of course followed the
improvements in computer capabilities and facilities. \footnote{The author
feels compelled to remind of the difficulties in 1980's when he and others
had to 'escape' to Europe where such facilities were more readily
available}
These have in particular benefitted the
Monte-Carlo and shell-model (in particular the no-core shell-model) calculations.

In astrophysics and heavy-ion collision problems one deals with nuclear matter
with densities well beyond those of the nuclear saturation point, that are not
reachable by Brueckner or similar methods. There are many efforts of extrapolation
by compressibility calculations with these methods.
The Rho-Brown scaling procedure is one approach to this problem.\cite{rho04}
There is evidence that 3-body forces are important in nuclear structure
calculations. At higher densities 4-body (and higher) forces are likely to come
into play. Brueckner theory, as usually practiced, includes medium-effects up to 
3-body terms. At higher densities higher order terms (diagrams) become increasingly
important. This is a challenging problem for theorists.

\end{document}